\documentclass[letterpaper,english,preprint, aps]{revtex4-1}
\usepackage[T1]{fontenc}
\usepackage[latin9]{inputenc}
\usepackage{amsmath}
\usepackage{amssymb}
\usepackage{graphicx}
\usepackage{setspace}

\makeatletter

\pdfpageheight\paperheight
\pdfpagewidth\paperwidth

\makeatother

\usepackage{babel}
\begin{document}
\preprint{BROWN-HET-1773}
\title{Notes on Scrambling in Conformal Field Theory}
\author{Chang Liu}
\email{chang_liu3@brown.edu}
\author{David A. Lowe}
\email{lowe@brown.edu}
\affiliation{Physics Department, Brown University, Providence, RI, 02912, USA}
\begin{abstract}
The onset of quantum chaos in quantum field theory may be studied
using out-of-time-order correlators at finite temperature. Recent
work argued that a timescale logarithmic in the central charge emerged
in the context of two-dimensional conformal field theories, provided
the intermediate channel was dominated by the Virasoro identity block.
This suggests a wide class of conformal field theories exhibit a version
of fast scrambling. In the present work we study this idea in more
detail. We begin by clarifying to what extent correlators of wavepackets
built out of superpositions of primary operators may be used to quantify
quantum scrambling. Subject to certain caveats, these results concur
with previous work. We then go on to study the contribution of intermediate
states beyond the Virasoro identity block. We find that at late times,
time-ordered correlators exhibit a familiar decoupling theorem, suppressing
the contribution of higher dimension operators. However this is no
longer true of the out-of-time-order correlators relevant for the
discussion of quantum chaos. We compute the contributions of these
conformal blocks to the relevant correlators, and find they are able
to dominate in many interesting limits. Interpreting these results
in the context of holographic models of quantum gravity, sheds new
light on the black hole information problem by exhibiting a class
of correlators where bulk effective field theory does not predict
its own demise.
\end{abstract}
\maketitle

\section{Introduction}

It has been suggested that quantum theories of gravity exhibit a property
known as fast scrambling, where a generic quantum state exhibits global
thermalization in a timescale that is logarithmic in the system size
\citep{Sekino:2008he}. It is interesting to explore this idea in
the context of holographic theories of gravity dual to conformal field
theories, where one may try to extract constraints on the class of
conformal field theories with gravity duals.

One simple way to to quantify this notion of scrambling is to consider
the norm (or equivalently the square) of the commutator of a pair
of Hermitian operators $V$ and $W$ at different times. For the purposes
of the present paper, we will also consider the system at finite temperature,
with inverse temperature $\beta$. This leads to a relation with out-of-time-order
correlators 

\begin{spacing}{1.3}
\begin{align}
-\langle[V(0),W(t)]^{2}\rangle_{\beta} & =\langle V(0)W(t)W(t)V(0)\rangle_{\beta}+\langle W(t)V(0)V(0)W(t)\rangle_{\beta}\nonumber \\
 & \quad-\langle W(t)V(0)W(t)V(0)\rangle_{\beta}-\langle V(0)W(t)V(0)W(t)\rangle_{\beta}\,.\label{eq:commutator}
\end{align}
For sufficiently late times, the first two terms are simply the time-independent
disconnected diagram $\langle WW\rangle_{\beta}\langle VV\rangle_{\beta}$,
while the last two terms are genuine out-of-time-order correlators.
For the 2d conformal field theories of interest here, these correlators
may be computed by continuing the Euclidean four-point function through
the second Riemann sheet \citep{PhysRevLett.115.131603}, as we describe
in detail later. These terms vary as a function of $t$, unlike the
disconnected terms, and from them a scrambling timescale may be extracted.
In the following section we describe in more detail the dependence
of this timescale on the chosen operators. Briefly, one wishes to
choose operators that exhibit the longest scrambling timescale, so
one may use this commutator computation as a proxy for asking that
the longest timescale a generic state scrambles. There may of course
exist special choices of operators with much shorter scrambling times,
and likewise special choices with much longer times, such as those
that commute with the Hamiltonian.
\end{spacing}

In order to study these out-of-time-order correlators at finite temperature
in conformal field theory we will begin with the Euclidean theory
on $S^{1}\times\mathbb{R}$. The correlators in this theory may be
obtained by a conformal mapping from the complex plane. The circle
direction is to be periodically identified with period $\beta$ and
corresponds to the imaginary time direction. The spatial direction
is then necessarily of infinite extent. For the purposes of the present
paper we will study four-point correlators of primary operators, as
well as correlators of wavepackets of such operators. Four-point functions
of primaries are expressed in the so-called conformal blocks of the
theory. In general, these conformal blocks are not known beyond infinite
series expansions. However there has been much progress in the literature
on obtaining asymptotic expansions of these conformal blocks in a
variety of limits, and we will make extensive use of these results
in the following \citep{Fitzpatrick:2014vua}.

In holographic theories, the graviton mode is dual to the stress energy
tensor of the CFT, which in turn is a Virasoro descendant of the identity
operator. Long distance bulk physics should be dominated by the propagation
of this mode, so the limit where the identity block dominates the
conformal block is of particular interest. Assuming this intermediate
Verma module dominates the conformal block of the four-point function
\citep{PhysRevLett.115.131603} (as well as assuming large central
charge and large external conformal weight $h_{w}$) obtained a scrambling
time logarithmic in the central charge $c$ of the CFT
\begin{equation}
t_{*}=\frac{\beta}{2\pi}\log\frac{c}{h_{w}}\label{eq:rstime}
\end{equation}
suggesting (at least if the result can be continued to values $h_{w}$
of order 1) that conformal field theories exhibit a version of fast
scrambling.

\begin{spacing}{1.3}
In this paper we will study this problem in more detail. One immediate
issue is that primary operators on their own do not exhibit the timescale
\eqref{eq:rstime}, but rather a thermalization timescale of order
$\beta$ or less. However the class of states obtained by acting on
the thermal state with a primary is not necessarily a good representative
of a generic state, so this is not an immediate contradiction. To
proceed we fold the primary operators into wavepackets, and consider
optimizing the shape of the wavepacket to obtain the longest thermalization
time. When this is done, we find a timescale resembling \eqref{eq:rstime}
does indeed emerge. Next we examine the contribution of Verma modules
with higher conformal weights to the four-point function. While we
find the time-ordered four-point functions respect the familiar late-time
decoupling theorems, and can be ignored with respect to the identity
block, this is no longer true of the out-of-time-order correlators
needed to compute \eqref{eq:commutator}. We compute the contributions
of these higher intermediate states, and find these can indeed dominate
the commutator even when all the time-ordered correlators have a sensible
holographic description in terms of bulk low energy effective field
theory. This implies that many of the bulk observables, defined over
finite ranges of time, that one might use to probe the black hole
information problem, are not accessible using low energy effective
field theory. In this sense effective field theory does not predict
its own demise.
\end{spacing}

\section{Scrambling and CFT Correlators}

\begin{figure}
\centering\includegraphics[width=0.41\textwidth]{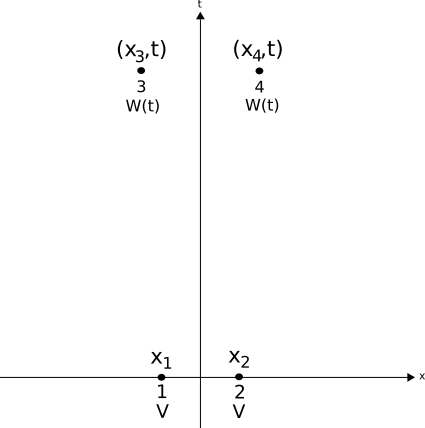} \caption{\label{fig:localops}The configuration of four pair-wise identical
local operators, themselves separated by large $t$.}
\end{figure}

\begin{spacing}{1.3}
We consider a thermal system described by a conformal field theory
living on a spatial real line $x$ with imaginary time $-it$ periodically
identified with period $\beta$. We can map this spatially infinite
thermal system to a CFT defined on the complex plane $z$ via the
exponential map
\[
z=\exp\left(\frac{2\pi}{\beta}(x+t)\right)\,.
\]
We are interested in computing the 4-point functions that appear in
\eqref{eq:commutator} so to this end we consider four pair-wise local
operators, inserted at distinct spatial positions as in fig.~\ref{fig:localops}.
We therefore have, after conformal mapping 
\[
\begin{aligned}z_{1} & =e^{\frac{2\pi}{\beta}x_{1}}\\
z_{2} & =e^{\frac{2\pi}{\beta}x_{2}}\\
z_{3} & =e^{\frac{2\pi}{\beta}(x_{3}+t)}\\
z_{4} & =e^{\frac{2\pi}{\beta}(x_{4}+t)}
\end{aligned}
\]
where we are interested in the limit $x_{1}\to x_{2},\,x_{3}\to x_{4}$
to reproduce the desired commutator.

The spacetime dependence of the conformal blocks appearing in the
4-point function will only depend on the cross-ratio $z=z_{12}z_{34}/z_{13}z_{24}$
(and $\bar{z}$) which is easily shown to be 
\[
z=\frac{\sinh\left(\frac{\pi}{\beta}(x_{1}-x_{2})\right)\sinh\left(\frac{\pi}{\beta}(x_{3}-x_{4})\right)}{\sinh\left(\frac{\pi}{\beta}(t-x_{1}+x_{3})\right)\sinh\left(\frac{\pi}{\beta}(t-x_{2}+x_{4})\right)}\,.
\]
As discussed in appendix \ref{sec:Correlators-and-Conformal} we rescale
the 4-point function by the coincident 2-point functions, to scale
out the operator norm. The rescaled correlators then depend only on
the cross-ratios as in \eqref{eq:normcorr}.
\end{spacing}

As an example, let us consider the identity conformal block in a large
$c$ limit, where the $V$ and $W$ operators have conformal weights
$h_{v}$ and $h_{w}$ respectively. The large $c$ limit is to be
taken with $h_{w}/c$ fixed, and $h_{v}\ll c$ fixed. The conformal
block $\mathcal{F}(z)$ in this limit is computed in \citep{Fitzpatrick:2014vua,Fitzpatrick:2015zha}
\begin{equation}
z^{2h_{v}}\mathcal{F}(z)\approx\left[\frac{z\alpha_{w}(1-z)^{(\alpha_{w}-1)/2}}{1-(1-z)^{\alpha_{w}}}\right]^{2h_{v}}\,,\label{eq:prinblock}
\end{equation}
with $\alpha_{w}=\sqrt{1-24h_{w}/c}$. The real-time out-of-time-order
correlator is obtained by continuing this block to the second Riemann
sheet as described in \citep{PhysRevLett.115.131603} and the leading
contribution to the rescaled commutator is 
\begin{equation}
z^{2h_{v}}\mathcal{F}(z)\approx\left[\frac{e^{-\pi i(\alpha_{w}-1)}z\alpha_{w}(1-z)^{(\alpha_{w}-1)/2}}{1-e^{-2\pi i\alpha_{w}}(1-z)^{\alpha_{w}}}\right]^{2h_{v}}\sim\left(\frac{1}{1-\frac{24\pi ih_{w}}{cz}}\right)^{2h_{v}}\,.\label{eq:identityblock}
\end{equation}
Let us take a limit where $\epsilon_{12}=x_{1}-x_{2}$ and $\epsilon_{34}=x_{3}-x_{4}$
are much smaller than $\beta$, and without loss of generality set
$x_{1}=0$. The cross-ratio is then approximately
\[
z\approx\frac{\pi^{2}}{\beta^{2}}\frac{\epsilon_{12}\epsilon_{34}}{\sinh^{2}\left(\frac{\pi}{\beta}\left(t+x_{3}\right)\right)}
\]
provided we stay away from light-like separations where $x_{3}\to-t$.
As we see the conformal block on the second sheet has a simple limit
as $\epsilon_{12}$ and $\epsilon_{34}\to0$, when $z\to0$, corresponding
to the actual computation of the commutator
\begin{equation}
z^{2h_{v}}\mathcal{F}(z)\approx\left(\frac{cz}{24\pi ih_{w}}\right)^{2h_{v}}\,.\label{eq:latetimeblock}
\end{equation}
The exponential decay of this quantity indicates the commutator between
$V$ and $W$ becomes large after a time of order 
\begin{equation}
t=\frac{\beta}{4\pi h_{v}}\label{eq:primarytime}
\end{equation}
showing rapid thermalization of primary operators on a timescale much
shorter than \eqref{eq:rstime}. 

However the interesting physical question is whether generic states
exhibit some notion of quantum scrambling on a longer timescale. To
explore this question in the current context of CFT 4-point functions,
we can then try to build more generic deformations of the thermal
density matrix by acting with primary operators folded into wavepackets
with some characteristic spatial size $L$. Computing the 4-point
function of these wavepackets, one can attempt to vary $L$ to maximize
the convoluted amplitude, then ask what thermalization timescale emerges.

Concretely, we convolute the function \eqref{eq:identityblock} with
spatial Gaussian wavepackets with width $L$. We will choose $t,L$
and the $x_{i}$ such that light-like singularities in $z$ are avoided.
In this regime, the resulting integral will be dominated by a saddle
point value of $z$, and the convoluted (rescaled) conformal block
may then be well approximated by simply substituting this value into
\eqref{eq:identityblock}. Given the simple form of \eqref{eq:identityblock},
with a cusp at $z=1$, the optimal value for $L$ will be the one
that makes $z$ approach $1$.

For simplicity let us set $x_{1}+x_{2}=x_{3}+x_{4}=0$, and we will
build Gaussian wavepackets in the variables $x_{1}-x_{2}=\ell_{v}$
and $x_{3}-x_{4}=\ell_{w}$. To fix $L$ in terms of $z$, one is
therefore interested in the convolution
\begin{equation}
z(t,L)=\frac{4}{\pi L^{2}}\int_{0}^{\infty}dl_{v}dl_{w}e^{-(l_{v}^{2}+l_{w}^{2})/L^{2}}\frac{\sinh\left(\frac{\pi}{\beta}l_{v}\right)\sinh\left(\frac{\pi}{\beta}l_{w}\right)}{\sinh\left(\frac{\pi}{\beta}\left(t-\frac{l_{v}}{2}+\frac{l_{w}}{2}\right)\right)\sinh\left(\frac{\pi}{\beta}\left(t+\frac{l_{v}}{2}-\frac{l_{w}}{2}\right)\right)}\,.\label{eq:crossratiosmear}
\end{equation}
This formula is justified because the exponential variation of $z$
with $l_{v},l_{w}$ is much more rapid than power law variation of
the conformal block with $z$, so analyzing the convolution of $z$
alone is sufficient to determine $l_{v}$ and $l_{w}$ and subsequently
$L$. The integrand has light-like poles, however for suitable values
of $t$ and $L$ these contributions to the smeared conformal block
can be made negligible. In this limit, the integrand can be well-approximated
by simply
\[
z(t,L)\approx\frac{4}{\pi L^{2}}\int_{0}^{\infty}dl_{v}dl_{w}e^{-(l_{v}^{2}+l_{w}^{2})/L^{2}}\frac{2\sinh\left(\frac{\pi}{\beta}l_{v}\right)\sinh\left(\frac{\pi}{\beta}l_{w}\right)}{\cosh\left(\frac{2\pi}{\beta}t\right)}\,.
\]
This has saddle points when
\[
l_{v}\tanh\left(\frac{l_{v}\pi}{\beta}\right)=\frac{\pi L^{2}}{2\beta}
\]
and likewise for $l_{w}$. The positive solutions are to be taken
corresponding to the limits of integration in \eqref{eq:crossratiosmear}.
If we then ask that the resulting amplitude \eqref{eq:identityblock}
is maximized in magnitude, we find that we must choose $L\sim\beta$
near $t=0$. We choose not to change the shape of the wavepackets
at time increases, and impose this condition for all values of $t$.
At the end we find the optimal value of $z$ is
\begin{equation}
z_{sad}=\mathrm{sech}\left(\frac{2\pi}{\beta}t\right)\label{eq:zsad}
\end{equation}
up to constant factors of order $1$. 

\begin{figure}
\includegraphics[width=0.45\textwidth]{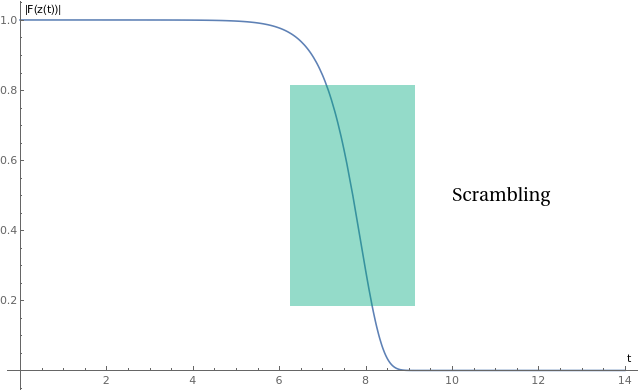}\qquad{}\includegraphics[width=0.45\textwidth]{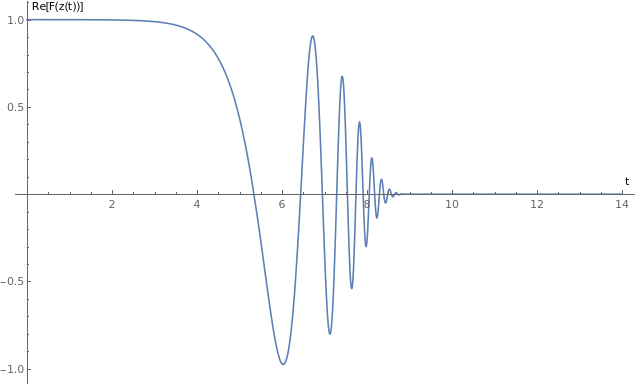}
\caption{\label{fig:scrambling-plot}Plot of function $|z^{2h_{v}}\mathcal{F}(z)|=|F(z(t))|=1/\left|1-12\pi ih_{w}\exp\left(2\pi/\beta\left(t-\log c-x\right)\right)\right|^{2h_{v}}$
where $c=10^{7}$, $h_{v}=100$, $h_{w}=10$, $\beta=2\pi$ and $x=0$.
Here $t_{*}=7.7$ according to \eqref{eq:scrambletime}. In the right
panel, a plot of $\mathrm{Re}\,F(z(t))$ is shown.}
\end{figure}

Let us now return to the example of the identity conformal block continued
to the second Riemann sheet as considered in \citep{PhysRevLett.115.131603}.
In this case, the saddle point approximation to the (rescaled) convoluted
block function is for sufficiently late times
\begin{equation}
z^{2h_{v}}\mathcal{F}(z)\approx\left(\frac{1}{1-\frac{12\pi ih_{w}}{c}e^{\frac{2\pi}{\beta}(t-x)}}\right)^{2h_{v}}\label{eq:rscorrel}
\end{equation}
where we have restored dependence on the spatial separation $x$ of
the centers of the wavepackets, and inserted the saddle point approximation
value for $z$ \eqref{eq:zsad} for $t\gg\beta$. It is helpful to
plot this for sample parameters as in fig. \ref{fig:scrambling-plot}.
As $t-x$ increases from $0$ to 
\begin{equation}
t_{*}=\frac{\beta}{2\pi}\log\frac{c\sqrt{\log2}}{12\pi h_{v}^{1/2}h_{w}}\label{eq:scrambletime}
\end{equation}
the conformal block decreases in magnitude by a factor of about $1/2$.
This thermalization time may be viewed as a proxy for the true scrambling
time of the system, and shows the distinctive appearance of the logarithm
of the system size. The formula is valid for $0<h_{v}\ll c$, but
ideally one would want to argue this formula continues to hold as
$h_{w}$ becomes of order $1$. Unfortunately it is not yet possible
to prove this. We note fig. \ref{fig:scrambling-plot} also shows
in the late-time limit the asymptotic form \eqref{eq:latetimeblock}
is applicable and the timescale for variation is the much shorter
time \eqref{eq:primarytime}.

The correlator of the wavepackets is given by \eqref{eq:rscorrel}
provided one steers clear of the light-cone singularities in \eqref{eq:crossratiosmear}
which render the approximation \eqref{eq:zsad} invalid. This is a
signature that even the wavepackets of primaries are not ideal representatives
of a generic state, and retain regions of spacetime where thermalization
has not yet occurred, outside the light-cone of the wavepacket. Nevertheless
for the present purposes, the reduced state inside the light-cone
appears to be well-thermalized according to the correlators, so this
procedure should yield a good measure of the global scrambling time.
Again it remains to be seen whether \eqref{eq:scrambletime} holds
in the case of most physical interest where $h_{w}$ is of order 1.

\begin{spacing}{1.3}

\end{spacing}

\section{Higher Weight Intermediate States }

\begin{spacing}{1.3}
We now turn our attention to the contribution of higher weight intermediate
states to the out-of-time order correlators, and will find the surprising
result that these may dominate over the identity block in the late-time
limit. Again we will assume we are taking $c\gg1$ with $h_{w}/c$
fixed and $h_{v}\ll c$ fixed. In addition we will generalize from
the identity block to an intermediate channel with conformal weight
$h_{p}$ fixed as $c\to\infty$.
\end{spacing}

Our starting point is the formula for the conformal block at next-to-leading
order in this large $c$ expansion of \citep{Fitzpatrick:2015zha}

\[
\mathcal{F}(z)=\mathcal{F}_{0}(z)\left(\frac{1-(1-z)^{\alpha_{w}}}{\alpha_{w}}\right)^{h_{p}}{}_{2}F_{1}\left(h_{p},h_{p},2h_{p},1-(1-z)^{\alpha_{w}}\right)
\]
where $_{2}F_{1}(\alpha,\beta;\gamma;z)$ is the Gauss hypergeometric
function. To continue this expression to the second sheet we use the
hypergeometric function identity \citep{hypergeomexp}
\[
\frac{\Gamma(h)^{2}}{\Gamma(2h)}{}_{2}F_{1}(h,h;2h;w)=\left(\sum_{k=0}^{\infty}\frac{2\left(h\right)_{k}^{2}\left(\psi(k+1)-\psi(h+k)\right)}{k!^{2}}\ensuremath{(1-w)^{k}}\right)-\log(1-w)\,_{2}F_{1}(h,h,1;1-w)
\]
 valid for $|1-w|<1$, where $\left(h\right)_{k}$ is the Pochhammer
symbol, and $\psi(a)$ is the digamma function. Continuing to the
second sheet we then obtain
\begin{align}
\mathcal{F}_{II}(z) & =\mathcal{F}_{0,II}(z)\left(\frac{1-e^{-i2\pi\alpha_{w}}(1-z)^{\alpha_{w}}}{\alpha_{w}}\right)^{h_{p}}\left(_{2}F_{1}\left(h_{p},h_{p},2h_{p},1-e^{-i2\pi\alpha_{w}}(1-z)^{\alpha_{w}}\right)\right.\nonumber \\
+ & \left.2\pi i\alpha_{w}\frac{\Gamma(2h_{p})}{\Gamma(h_{p})^{2}}\,_{2}F_{1}(h_{p},h_{p},1;e^{-i2\pi\alpha_{w}}\left(1-z\right)^{\alpha_{w}})\right)\,.\label{eq:blocktwo}
\end{align}
Expanding for small $h_{w}/c$ and $z\ll1$ leads to
\[
\mathcal{F}_{II}(z)\sim\mathcal{F}_{0,II}(z)\left(\frac{z-\frac{\pi ih_{w}}{6c}}{\alpha_{w}}\right)^{h_{p}}\left(1+i\tan\left(\pi h_{p}\right)-2\pi^{2}iz^{1-2h_{p}}\frac{\Gamma(2h_{p})}{\Gamma(2-2h_{p})\Gamma(h_{p})^{4}\sin\left(2\pi h_{p}\right)}\right)\,.
\]
This ends up being dominated by the last term in the third factor,
and in fact grows at late times. Even at early times ($z$ near 1)
the last term in \eqref{eq:blocktwo} dominates over the other term
in the third factor for $h_{p}>1$. The second factor in \eqref{eq:blocktwo}
rapidly approaches a constant much smaller than 1. 

The upshot is the identity block dominates for a finite period of
time, however after 
\[
t_{*}\approx\frac{\beta}{4\pi}\log\left(\frac{c}{h_{w}}\right)
\]
the higher weight intermediate states take over. This late time sum
over intermediate states apparently diverges when considered term
by term. This would lead one to conclude the commutator grows initially
while dominated by the identity block, but then may again decrease
at later times, indicating a lack of true scrambling in the conformal
field theory.

One possible way to avoid this conclusion is to demand an infinite
tower of higher weight intermediate primaries, such that the apparently
divergent sum might be resummed to a finite answer. However in the
following section we find contributions for $h_{p}\gg c$ are actually
suppressed. We conclude that even a sparse spectrum of intermediate
primaries with weights $1<h_{p}\ll c$ are sufficient to destroy or
drastically modify the onset of quantum chaos. In light of our previous
discussion, this may simply mean such smeared primaries are still
not good representatives of generic states, and instead one would
need to consider commutators of much more general operators to see
the correct timescale for global thermalization, or quantum scrambling.
Alternatively, it may happen that only operators dual to black hole
states efficiently scramble, and these must be reflected in a choice
of external operators that do not couple at all (or only very weakly)
to higher weight primaries, such that the identity block may dominate
the out-of-time order correlators.

For conformal field theories with holographic anti-de Sitter gravity
duals, the implication of the higher intermediate channels is that
the bulk effective field theory breaks down when it is used to compute
out-of-time-ordered correlators at finite time. On the other hand,
there is no indication of such a breakdown when time-ordered CFT correlators
are computed (see also \citep{Fitzpatrick:2016ive,Fitzpatrick:2016mjq}),
which correspond to the boundary $S$-matrix of the bulk theory. To
see this we simply note that as higher dimensional operators in CFT$_{2}$
correspond to interactions of increasing mass scale in AdS$_{3}$,
domination of all intermediate channels with dimension $h_{p}\geq1$
means that there would be a dual set of an infinite sequence of interactions
in the gravitation theory in AdS$_{3}$. If these high scale interactions
affect the infrared physics of the theory, then the standard decoupling
theorems of effective field theory such \citep{PhysRevD.11.2856}
break down. 

Now the usual measurements we perform can be well-approximated by
transition amplitudes, built out of time-ordered correlators which
may be computed as within effective field theory. It is only the particular
set of observables corresponding to out-of-time-order correlators,
or norms of commutators that exhibit this peculiar behavior. For the
black hole information problem this would seem to imply that contrary
to expectations, commutators that measure limits on the causal propagation
of information are indeed observables sensitive to the ultra-violet
structure of the theory, as long hinted at in perturbative string
theory computations \citep{Lowe:1995pu,Lowe:1995ac}.

\section{Intermediate channels with $h_{p}\gg c$}

\begin{spacing}{1.3}
So far we have only considered intermediate channels with fixed $h_{p}\ll c$.
It is also instructive to perform the same analysis for intermediate
channels with $h_{p}\gg c$ where the limit is $h_{p}\to\infty$ with
$c/h_{p}$, $h_{v}/h_{p}$ and $h_{w}/h_{p}$ fixed and small. For
this we consider equation (16) in \citep{Zam87}, 
\begin{equation}
\mathcal{F}(z)=\left(16q\right)^{h_{p}-\frac{c}{24}}z^{\frac{c}{24}-2h_{v}}(1-z)^{\frac{c}{24}-(h_{v}+h_{w})}\theta_{3}(q)^{\frac{c}{2}-8(h_{v}+h_{w})}H(c,h_{p},h_{i},q)\label{eq:eq16zem87}
\end{equation}
where the nome $q=e^{i\pi\tau}$ is related to the cross-ratio $z$
by 
\[
\tau=i\frac{K'(z)}{K(z)}=i\frac{K(1-z)}{K(z)}
\]
where $K(z)$ is the complete elliptic integral with parameter\footnote{We clarify that in most mathematical literature, the complete elliptic
integral $K$ is defined with the modulus $k$ as the argument. Our
$z$ is related to $k$ through $z=k^{2}$. It is also common for
many mathematicians to use the symbol $m$ for our $z$.} $z$. Here $H$ is a function that is $1+O(1/h_{p})$ and 
\begin{equation}
\theta_{3}(q)=\sum_{n=-\infty}^{\infty}q^{n^{2}}\,.\label{eq:thetathree}
\end{equation}

Eq.~(\ref{eq:eq16zem87}) has a branch cut at $z=1$ from the $1-z$
factor which will lead to the same analytic behavior for the intermediate
case $h_{p}\ll c$, which we have previously considered. To see this
we expand the nome $q$ around $z=0$ to obtain
\[
q=e^{i\pi\tau}=\frac{z}{16}+\frac{z^{2}}{32}+\cdots\,.
\]
As $\theta_{3}(q)$ is regular near $q=0$, we see that on the principal
sheet $\mathcal{F}(z)$ goes to zero as $z\to0$. Therefore the heavy
intermediate channels are perfectly suppressed on the first Riemann
sheet. Crossing the branch cut $z=1$ from above, the complete elliptic
function $K(z)$ picks up an additional imaginary part \citep{Bogner:2017vim}:
\[
\lim_{\epsilon\to0^{+}}K(z+i\epsilon)=K(z)+2iK(1-z)\,.
\]
Analyticity implies that on the second Riemann sheet the nome is now
\[
q=\exp\left[-\frac{\pi K(1-z)}{K(z)+2iK(1-z)}\right]=\exp\left[-\frac{\pi}{\frac{K(z)}{K(1-z)}+2i}\right]\,.
\]
To expand this expression near $z=0$, we use
\[
\frac{K(z)}{K(1-z)}\approx\frac{\pi}{4\log2-\log z}+\mathcal{O}\left(\frac{z}{\log^{2}z}\right)
\]
so that
\begin{equation}
q\approx e^{\frac{i\pi}{2}+\frac{\pi^{2}}{4\log z}}\,.\label{eq:nomeexpan}
\end{equation}
We then need to expand \eqref{eq:thetathree} near $q=i$. The expansion
near $q=1$ is 
\[
\left|\theta_{3}(q)\right|\approx\left|\frac{\sqrt{\pi}}{\sqrt{1-q}}\right|
\]
but we can obtain the expansion near $q=i$ by using the relation
\[
\left|\theta_{3}(q)\right|=\left|\frac{\sqrt{\pi}}{\sqrt{\log q}}\theta_{3}\left(e^{\frac{\pi^{2}}{\log q}}\right)\right|
\]
and substituting in \eqref{eq:nomeexpan} to give $\theta_{3}(q)$
near $q=i$ as 
\begin{equation}
\left|\theta_{3}(q)\right|\approx\left|\frac{\sqrt{-2\log z}}{\sqrt{\pi}}\right|\,.\label{eq:nomeneari}
\end{equation}
Assembling the various factors, we find again a dramatic enhancement
of the higher weight channel on the second Riemann sheet arising from
the behavior \eqref{eq:nomeneari}, compared to the behavior on the
principal sheet. However when we compare to the $h_{p}=0$ expression
of the previous section, the $z^{c/24}$ factor of \eqref{eq:eq16zem87}
dominates for small $z$ so we conclude they do not dominate versus
the identity channel (again modulo restrictions on the operator couplings
$C_{p}$ of \eqref{eq:blockdef}).
\end{spacing}

\section{Conclusions}

\begin{spacing}{1.3}
In this paper we discussed the issue of smearing local operators in
a thermal CFT and its connection with quantum scrambling. We pointed
out that the correct scrambling time should be identified with operators
that maximize the timescale of variation of the out-of-time ordered
correlator, which may occur well before the asymptotic late-time limit.
We then examined a somewhat independent issue, that the higher intermediate
states with $0<h_{p}\ll c$ can have large out-of-time ordered correlators.
We discussed the implications of this statement, which is that in
the AdS$_{3}$ gravity dual the UV dynamics and IR dynamics is no
longer decoupled when these observables are computed. This lack of
decoupling appears even when the usual time-ordered correlators, or
transition amplitudes satisfy the standard decoupling lore. When applied
to scattering in $AdS_{3}$ black hole backgrounds this implies that
the commutators that lead one to conclude information is lost semiclassically,
are in fact not computable without a full specification of the ultraviolet
physics of the theory. The ordinary bulk effective field theory does
not predict its own demise when computing these observables.
\end{spacing}

As for the appearance of a scrambling time of the form \eqref{eq:rstime}
we have found a variant of this expression \eqref{eq:scrambletime},
valid when the identity block dominates. The expression involves a
term of the form $\beta/2\pi\log c$, but other significant terms
are also present. If other intermediate primaries appear, with conformal
weights fixed in a large $c$ limit, they will dominate the late-time
behavior and may completely spoil thermalization. It will be very
interesting to extend the range of validity of these expressions to
determine whether there exist a class of 2d conformal field theories
that may be viewed as fast scramblers at finite temperature. 
\begin{acknowledgments}
D.L. thanks S. Hellerman and E. Perlmutter for helpful comments. We
thank B. Stoica, A. Rolph and H. Hampapura for help in checking the
semiclassical exponentiation formula for conformal blocks. We thank
R. Fan for helpful comments on an earlier draft and note the closely
related work in the context of $c<1$ unitary minimal models \citep{Fan:2018ddo}.
D.L. is supported in part by DOE grant DE-SC0010010. This work was
completed at the Aspen Center for Physics, which is supported by National
Science Foundation grant PHY-1607611.
\end{acknowledgments}

\appendix*

\section{\label{sec:Correlators-and-Conformal}Correlators and Conformal Blocks}

Conformal blocks are usually written in terms of 4-point functions
after a global $SL(2,C)$ conformal transformation has sent generic
points in the complex plane to the values $0,z,1,\infty$. Here we
briefly unpack the relation between these conformal blocks and 4-point
functions for general $z_{i}$.

A canonical form for the 4-point function at general $z_{i}$ in the
complex plane is \citep{Ginsparg:1988ui}
\begin{equation}
\left\langle \prod_{i=1}^{4}\mathcal{O}_{i}(z_{i})\right\rangle =f(z,\bar{z})\prod_{i<j}z_{ij}^{-(h_{i}+h_{j})+h/3}\prod_{i<j}\bar{z}^{-(\bar{h}_{i}+\bar{h}_{j})+\bar{h}/3}\label{eq:canfour}
\end{equation}
where $z_{ij}=z_{i}-z_{j}$, the cross-ratio $z=z_{12}z_{34}/z_{13}z_{24}$
and $h=\sum_{i}h_{i}$. The conformal block on the other hand is usually
defined \citep{Belavin:1984vu} for the special choice $z_{i}=0,z,1,\infty$.
To define the correlator as the point $z_{4}$ moves to infinity we
must rescale by a factor of $z_{4}^{2h_{w}}$ 
\begin{equation}
\lim_{z_{4}\to\infty}z_{4}^{2h_{w}}\bar{z}_{4}^{2\bar{h_{v}}}\left\langle \prod_{i=1}^{4}\mathcal{O}_{i}(z_{i})\right\rangle =\sum_{p}C_{12p}C_{34p}\mathcal{F}(p,z)\bar{\mathcal{F}}(p,\bar{z})\,.\label{eq:blockdef}
\end{equation}
Comparing the two formulae yields
\begin{align}
\lim_{z_{4}\to\infty}z_{4}^{2h_{w}}\bar{z}_{4}^{2\bar{h_{v}}}\left\langle \prod_{i=1}^{4}\mathcal{O}_{i}(z_{i})\right\rangle \bigg|_{z_{1}=0,z_{3}=1,z_{2}=z} & =f(z,\bar{z})\left(1-z\right)^{h/3-h_{2}-h_{3}}z^{h/3-h_{1}-h_{2}}\nonumber \\
 & \times\left(1-\bar{z}\right)^{\bar{h}/3-\bar{h}_{2}-\bar{h}_{3}}\bar{z}^{\bar{h}/3-\bar{h}_{1}-\bar{h}_{2}}\nonumber \\
 & =\sum_{p}C_{12p}C_{34p}\mathcal{F}(p,z)\bar{\mathcal{F}}(p,\bar{z})\label{eq:blockcan}
\end{align}
and we see the canonical form of the 4-point function involves a nontrivial
rescaling of the conformal block by a function of the cross-ratio. 

Later when we study the commutator of two operators, $V$ and $W$
as a function of time, it will be convenient to factor out the norm
of the operators. To accomplish this we compute 
\begin{equation}
\frac{\left\langle V(z_{1})V(z_{2})W(z_{3})W(z_{4})\right\rangle }{\left\langle V(z_{1})V(z_{2})\right\rangle \left\langle W(z_{3})W(z_{4})\right\rangle }=z^{2h_{v}}\bar{z}^{2\bar{h}_{v}}\sum_{p}C_{12p}C_{34p}\mathcal{F}(p,z)\bar{\mathcal{F}}(p,\bar{z})\label{eq:normcorr}
\end{equation}
using \eqref{eq:canfour} and \eqref{eq:blockcan}. Now the expression
for general $z_{i}$  is a function only of the cross-ratios. Finally
we note that in performing a coordinate transformation to a different
coordinate system, each correlator of primaries transforms by
\[
\left\langle \prod_{i}\mathcal{O}(x_{i})\right\rangle =\prod_{i}\left(\frac{\partial z}{\partial x}\right)_{z=z_{i}}^{h_{i}}\left(\frac{\partial\bar{z}}{\partial\bar{x}}\right)_{\bar{z}=\bar{z}_{i}}^{\bar{h}_{i}}\left\langle \prod_{i}\mathcal{O}(z_{i})\right\rangle 
\]
and these factors cancel in the expression \eqref{eq:normcorr}.

\bibliographystyle{utphys}
\bibliography{cft_scrambling_ref}

\end{document}